\documentclass[prx,reprint,superscriptaddress,longbibliography]{revtex4-2}

\usepackage[dvipdfmx]{graphicx}
\usepackage{dcolumn}
\usepackage{bm}
\usepackage[normalem]{ulem}
\usepackage{braket}
\usepackage{float}
\usepackage{amsmath,amssymb,bm}
\usepackage{ascmac}
\usepackage{array}
\usepackage{ulem}
\usepackage{hyperref}
\usepackage{booktabs}
\hypersetup{%
colorlinks=True,
linkcolor=blue, 
citecolor=blue,
urlcolor=blue
}

\usepackage{mathtools}
\usepackage{multirow}
\usepackage{placeins} 
\usepackage{appendix}
\usepackage{mleftright}
\mleftright 

\graphicspath{{figure/}}

\begin{document}

\title{Flux-Trapping Fluxonium Qubit}

\author{Kotaro~Hida}
\email{hida@qipe.t.u-tokyo.ac.jp}
\affiliation{Department of Applied Physics, Graduate School of Engineering, The University of Tokyo, Bunkyo-ku, Tokyo 113-8656, Japan}
\author{Kohei~Matsuura}
\email{matsuura@qipe.t.u-tokyo.ac.jp}
\affiliation{Department of Applied Physics, Graduate School of Engineering, The University of Tokyo, Bunkyo-ku, Tokyo 113-8656, Japan}
\author{Shu~Watanabe}
\affiliation{Department of Applied Physics, Graduate School of Engineering, The University of Tokyo, Bunkyo-ku, Tokyo 113-8656, Japan}
\author{Yasunobu~Nakamura}
\affiliation{Department of Applied Physics, Graduate School of Engineering, The University of Tokyo, Bunkyo-ku, Tokyo 113-8656, Japan}
\affiliation{RIKEN Center for Quantum Computing (RQC), Wako, Saitama 351-0198, Japan}

\date{\today}

\begin{abstract}

In pursuit of superconducting quantum computing, fluxonium qubits have recently garnered attention for their large anharmonicity and high coherence at the sweet spot. Towards the large-scale integration of fluxonium qubits, a major obstacle is the need for precise external magnetic flux bias: To achieve high performance at its sweet spot, each qubit requires a DC bias line. However, such lines inductively coupled to the qubits bring in additional wiring overhead, crosstalk, heating, and decoherence, necessitating measures for mitigating the problems.
In this work, we propose a flux-trapping fluxonium qubit, which, by leveraging fluxoid quantization, enables the optimal phase biasing without using external magnetic flux control at the operating temperature. We introduce the design and working principle, and demonstrate the phase biasing achieved through fluxoid quantization.

\end{abstract}

\maketitle

\section{INTRODUCTION}
Towards practical quantum computation, the development of superconducting quantum circuits, especially transmon-based architectures, has been notably successful in the past decades~\cite{Wang2022, Acharya2023}. Transmon qubits~\cite{Koch2007} have been widely studied as a platform for superconducting quantum computers due to their design flexibility, ease of fabrication, robustness against noise, and model simplicity. However, their limited anharmonicity leads to inevitable frequency collisions on large chips~\cite{Hertzberg2021} and poses limitations on the gate speed.

Fluxonium qubits have emerged as an alternative platform for superconducting quantum computers~\cite{Manucharyan2009}. A fluxonium consists of a Josephson junction shunted by a capacitor and an inductor. It can be biased by an external magnetic flux threading the loop made of the Josephson junction and the inductor.
When biased at its flux frustration point with a minimum qubit frequency, often referred to as a sweet spot, a fluxonium becomes first-order insensitive to flux noise and has a small charge matrix element, resulting in a long coherence time. In fact, coherence times approaching and even exceeding 1~ms have been achieved in fluxoniums~\cite{Wang2024,Somoroff2023}. In addition, fluxoniums exhibit few-gigahertz anharmonicity at the sweet spots, which is beneficial for fast gate operations. Recently, fast and high-fidelity single-qubit gates~\cite{Rower2024} and two-qubit gates~\cite{Ding2023,Lin2024} have been demonstrated. These attractive properties have stimulated research on fluxonium quantum processors~\cite{Nguyen2022, Somoroff2024, Mazhorin2024, Lange2025, Zhao2025}.

On the other hand, the requirement for external magnetic flux control remains a major obstacle to their large-scale integration. To achieve the performance of fluxoniums at their sweet spots, precise control of the magnetic flux threading their loops is essential. This requirement necessitates DC bias lines coupled to each fluxonium and bias currents to generate precise magnetic fields. However, in the typical cryogenic environment where superconducting qubits are operated, such flux bias lines can be additional sources of decoherence due to their coupling to each qubit and the noise they introduce. 
Furthermore, wiring overhead, crosstalk between flux-tuning circuits, and passive and active heat load for the DC bias lines pose significant challenges to scaling up fluxonium-based quantum systems. 

\begin{figure*}[t]
\includegraphics{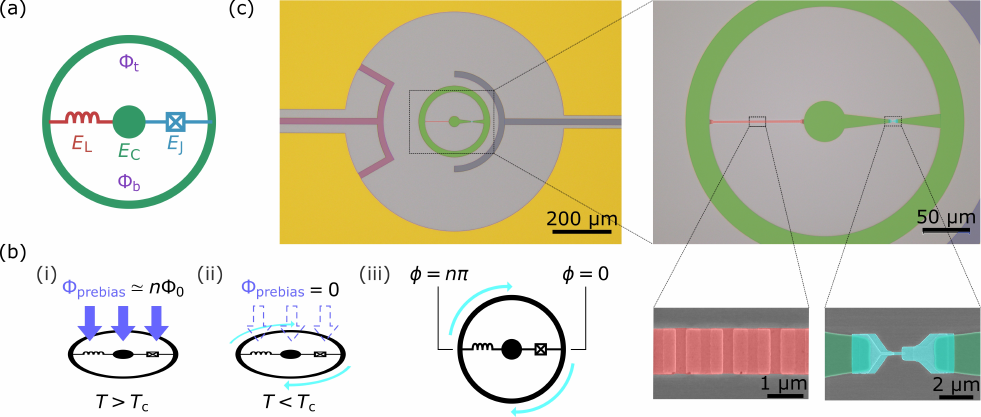}
\caption{Flux-trapping fluxonium.
(a) Schematic picture of a flux-trapping fluxonium. This circuit consists of a concentric capacitor~(green) shunted by a small Josephson junction~(light blue) and a superinductor made of a Josephson-junction array~(red), which respectively determine the qubit parameters $E_{\mathrm{C}}$, $E_{\mathrm{J}}$, and $E_{\mathrm{L}}$. The outer superconducting ring is for the flux trapping. The gradiometric structure allows additional phase biasing of the junction elements through the difference between the magnetic fluxes threading the top half, $\Phi_{\mathrm{t}}$, and the bottom half of the ring, $\Phi_{\mathrm{b}}$.
(b) Procedure of the flux-trapping operation detailed in the main text.
(c) False-colored images of the device. Superconducting structures including the ground plane~(yellow) are made of 100-nm-thick TiN film, except for the Josephson-junction electrodes made of Al, fabricated on a 300-$\mu$m-thick Si substrate~(gray). Left: Readout resonator~(blue) and on-chip bias line~(purple) coupled to the flux-trapping fluxonium. Right: Zoom-up image showing a concentric capacitor~(green), a small Josephson junction with the area of 200$\times$270~$\mathrm{n m^2}$~(blue), and a superinductor made of 91 serial Josephson junctions with each junction area of $\sim$1.5$\times$0.5~$\mathrm{\mu m^2}$~(red). 
}
\label{fig1}
\end{figure*}

Biasing fluxoniums at their sweet spots without using current bias lines can be a solution to mitigate the challenges. One approach is to incorporate a $\pi$-junction~\cite{Bulaevskii1977, Ioffe1999, Ryazanov2001}. Previous work~\cite{Kim2024} successfully demonstrated biasing a flux qubit at its sweet spot by utilizing a ferromagnetic $\pi$-junction with an error of $0.002\Phi_0$, where $\Phi_0 = h/2e$ is a flux quantum, corresponding to a single period of the phase bias. However, they also discussed the degradation of coherence properties due to the multi-layer structure in the $\pi$-junction.

Here, we focus on another approach, fluxoid quantization, to address the issue. Phase biasing by utilizing fluxoid quantization was first demonstrated in SQUID devices~\cite{Majer2002} and followed by experiments on flux qubits~\cite{Schwarz2013} and a fluxonium made of granular aluminum~\cite{Daria2022} at their sweet spots.
We further develop the scheme aiming at the integration of fluxoniums and characterize its performance. Our flux-trapping fluxonium has a thick and robust superconducting ring and a symmetric structure, enabling stable and accurate phase biasing at its sweet spot through fluxoid quantization. The flux-trapping fluxoniums achieve simultaneous phase biasing of multiple qubits at their sweet spots without requiring flux control at the operating temperature. Ideally, it eliminates dedicated bias lines for each fluxonium and simplifies flux biasing, leading to the mitigation of the problems related to the bias lines. 

This paper is organized as follows: In Sec.~\ref{sec:principle}, we introduce the architecture of the flux-trapping fluxonium and the flux-trapping operation to bias the qubit at its sweet spot. We characterize our qubits and demonstrate flux-trapping biasing in Sec.~\ref{sec:implement}, discuss the performance and the feasibility in Sec.~\ref{sec:discussion}, and summarize our work in Sec.~\ref{sec:conclusion}.

\section{Working principle} \label{sec:principle}

\subsection{Fluxoid quantization}\label{subsec:qunatization}

First, we briefly describe the formulation of fluxoid quantization.
Assuming a closed superconducting ring and its path $\mathcal{C}$, the periodic condition of the gauge-invariant phase $\phi$ leads to fluxoid quantization~\cite{London1948, Ginzburg1962}:
\begin{equation}\label{eq:fluxoid_quantization}
\frac{\Phi_0}{2\pi}\oint_{\mathcal{C}} \nabla \phi \cdot d\mathbf{r}=
\mu_0\oint_{\mathcal{C}} \lambda^2 \mathbf{J}_{\mathrm{s}} \cdot d\mathbf{r}+\int_{\mathcal{S}}\mathbf{B}\cdot d\mathbf{S}=n\Phi_0,
\end{equation}
where $\mu_0$ is the vacuum permeability, $\lambda$ is the magnetic penetration depth, $\mathbf{J}_{\mathrm{s}}$ is the superconducting current density, $\mathcal{S}$ is the area enclosed by $\mathcal{C}$, $\mathbf{B}$ is the magnetic field, and $n$ is an integer corresponding to the number of flux quanta trapped inside the ring. This formula indicates that the phase gradient along the ring is determined by $n$.

\subsection{Flux-trapping fluxonium qubit} \label{ssec:device}

\begin{figure*}[t]
\includegraphics{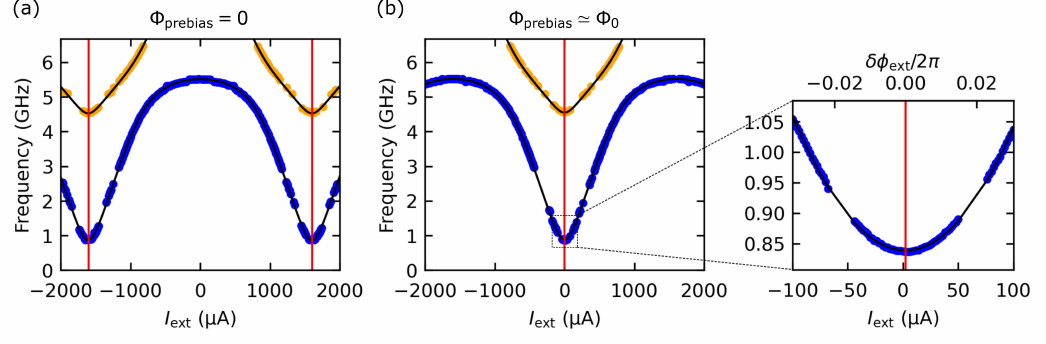}
\caption{Spectroscopy of the flux-trapping fluxonium. The blue~(orange) dots represent the measured frequencies of the first~(second) excited state of the qubit as a function of the current $I_{\mathrm{ext}}$ applied to the on-chip bias line. The black solid lines are fits to the fluxonium transition frequencies with $E_{\mathrm{J}}/h=3.54~\mathrm{GHz}$, $E_{\mathrm{C}}/h=1.32~\mathrm{GHz}$, and $E_{\mathrm{L}}/h=0.81~\mathrm{GHz}$. The red solid lines indicate the sweet spots.
(a) Excitation spectra with no flux quanta trapped in the ring.
(b) Left: Spectra when a single flux quantum was trapped. Right: High-resolution spectrum of the first excited state around the sweet spot.
$\delta\phi_{\mathrm{ext}}$ is the deviation from the sweet spot and defined as $\delta\phi_{\mathrm{ext}}\coloneqq\phi_{\mathrm{ext}}+\phi_{\mathrm{trap}}-\pi$; $\delta\phi_{\mathrm{ext}}=0$ corresponds to the sweet spot. 
}
\label{fig2}
\end{figure*}

We utilize the phase difference between two points on the ring to bias a qubit. 
First, we introduce the flux-trapping fluxonium. 
As shown in Fig.~\ref{fig1}(a), the qubit consists of a concentric capacitor shunted by a small Josephson junction and a superinductor made of a series of Josephson junctions. The outer superconducting ring is used for flux trapping and phase biasing. The qubit has a gradiometric structure with a mirror symmetry with respect to the center line along the small junction and the inductor. When there is no flux trapped in the ring, the qubit Hamiltonian can be described as
\begin{equation}\label{eq:Fluxonium_Hamiltonian}
\hat{\mathcal{H}}
=4E_{\mathrm{C}}\hat{n}^2-E_{\mathrm{J}}\cos{\hat{\phi}}+\frac{1}{2}E_{\mathrm{L}}(\hat{\phi}-\phi_{\mathrm{ext}})^2,
\end{equation}
where $\phi_{\mathrm{ext}} \coloneqq 2\pi(\Phi_{\mathrm{t}}-\Phi_{\mathrm{b}})/\Phi_0$ with $\Phi_{\mathrm{t(b)}}$ the external magnetic flux bias on the top (bottom) half of the ring.
Here, $\hat{n}$ and $\hat{\phi}$ are the charge and phase operators of the central node, where the phase origin $\phi = 0$ is defined at the right end of the ring~[see (iii) in Fig.~\ref{fig1}(b)]. $E_{\mathrm{C}}$, $E_{\mathrm{J}}$, and $E_{\mathrm{L}}$ are the charging, Josephson, and inductive energies, respectively. 

Next, we describe the flux-trapping operation to bias the qubit at its sweet spot without relying on $\phi_{\mathrm{ext}}$ induced by the gradient field. Figure~\ref{fig1}(b) illustrates the protocol: (i) At a temperature above the superconducting transition temperature $T_{\mathrm{c}}$, apply a magnetic flux $\Phi_{\mathrm{prebias}}$ threading the outer ring. (ii) Cool down the device through $T_{\mathrm{c}}$ while keeping the magnetic flux applied. 
Here we reconsider Eq.~(\ref{eq:fluxoid_quantization}). The first term of the middle part can be interpreted as a contribution from the kinetic inductance of the ring, and the second term as the sum of the contributions from the geometric inductance and the external magnetic field. We rewrite Eq.~(\ref{eq:fluxoid_quantization}) as
\begin{equation}\label{eq:fluxoid_quantization_2}
\Phi_{\mathrm{prebias}} + I_{\mathrm{s}}(L_{\mathrm{k}}+L_{\mathrm{g}})=n\Phi_0,
\end{equation}
where $I_{\mathrm{s}}$ is the supercurrent, and $L_{\mathrm{k}(\mathrm{g})}$ is the total kinetic~(geometric) inductance of the ring. To trap $n$ flux quanta in the ring, the inductive energy stored in the ring, 
\begin{equation}\label{eq:fluxoid_quantization_3}
E_{\mathrm{ring}} = \frac{1}{2} (L_{\mathrm{k}}+L_{\mathrm{g}})I^2_{\mathrm{s}} = \frac{(n\Phi_0-\Phi_{\mathrm{prebias}})^2}{2(L_{\mathrm{k}}+L_{\mathrm{g}})},
\end{equation}
has to be the lowest in terms of $n$. For that, the prebias flux $\Phi_\mathrm{prebias}/\Phi_0 \simeq n$ is chosen. At the qubit operation temperature $T \ll T_{\mathrm{c}}$, $n$ is not altered by any subsequent changes in the external magnetic flux, protected by the large phase-slip energy of the ring. Therefore, we can remove the applied magnetic flux. 
(iii)~As a result, the trapped flux gives a phase offset $\phi_{\mathrm{trap}}$ of $n\pi \pmod {2\pi}$ at the left end of the ring. The Hamiltonian is now written as
\begin{equation}\label{eq:Fluxonium_Hamiltonian2}
\hat{\mathcal{H}}
=4E_{\mathrm{C}}\hat{n}^2-E_{\mathrm{J}}\cos{\hat{\phi}}+\frac{1}{2}E_{\mathrm{L}}(\hat{\phi}-\phi_{\mathrm{trap}}-\phi_{\mathrm{ext}})^2.
\end{equation}
Most importantly, when $n$ is odd, $\phi_\mathrm{trap} = \pi \pmod {2\pi}$, and $\phi_{\mathrm{ext}} = 0$ gives the sweet spot of the fluxonium. In other words, we no longer need any external flux bias at its sweet spot at the qubit operation temperature.

Figure~\ref{fig1}(c) shows the design of our flux-trapping fluxonium, which has a 20-$\mathrm{\mu}$m-wide ring to secure the stability of flux trapping. The surrounding circuit structures around the qubit, including the pad of the readout resonator and the on-chip bias line, are also mirror symmetric about the symmetry axis of the gradiometric qubit to ensure the accuracy of the phase biasing. 
The on-chip bias line generates asymmetric magnetic flux to bias the gradiometric qubit without altering the total magnetic flux threading the ring. We use a superconducting coil made of NbTi surrounding the device package to generate a spatially homogeneous magnetic field for the flux-trapping operation~(See Appendix~\ref{sec:A} for the setup of our experiments).

\section{Implementation} \label{sec:implement}

\begin{figure*}[t]
\includegraphics{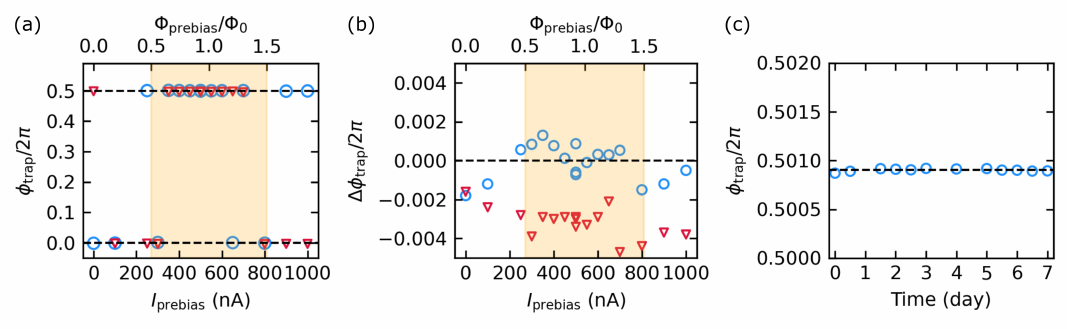}
\caption{Phase biasing due to flux trapping. Blue circles~(red triangles) represent the observed values for device 1~(device~2). $\Phi_{\mathrm{prebias}}$ on the top axes of (a) and (b) are the values of magnetic flux applied for the flux-trapping operation. 
The range of $0.5 \leq \Phi_{\mathrm{prebias}}/\Phi_0\leq1.5$, where a single flux quantum is expected to be trapped, is highlighted with an orange background color. 
(a) Phase bias $\phi_{\mathrm{trap}}$ due to flux trapping as a function of the applied prebias current $I_{\mathrm{prebias}}$ through the coil. The dashed black lines show the ideal bias point: $\phi_{\mathrm{trap}} = 0~\mathrm{or}~\pi$.
(b) Deviation of the phase bias from the target bias points, $\Delta \phi_{\mathrm{trap}}$. The dashed black line represents $\Delta\phi_{\mathrm{trap}} = 0$.
(c) Temporal stability of the phase bias in device 1. The black dashed line shows its average value of  $\phi_{\mathrm{trap}}/2\pi = 0.5+9.0\times10^{-4}$.
}
\label{fig3}
\end{figure*}

First, we performed spectroscopy of a flux-trapping fluxonium both with and without flux trapping. 
Figure~\ref{fig2}(a) shows the result of qubit spectroscopy without external magnetic flux during the superconducting transition, and Fig.~\ref{fig2}(b) shows the result with applying a current of $500~\mathrm{nA}$ through the coil for flux trapping.
From a calibration experiment using a SQUID-transmon qubit shown in Appendix~\ref{sec:C}, it was determined that a prebias current of $540~\mathrm{nA}$ through the coil generates a single flux quantum in the superconducting ring of our flux-trapping fluxonium. Thus, we expected the phase bias of $\phi_{\mathrm{trap}}=0$ in Fig.~\ref{fig2}(a) and $\phi_{\mathrm{trap}}=\pi$ in Fig.~\ref{fig2}(b).
In fact, as the red solid lines show, the sweet spots of Fig.~\ref{fig2}(a) were at $I_{\mathrm{ext}}\simeq \pm 1.6\times10^3~\mathrm{\mu A}$ and that of Fig.~\ref{fig2}(b) was at $I_{\mathrm{ext}}\simeq 0~\mathrm{\mu A}$. Thus, they corresponded to the cases with $\phi_{\mathrm{trap}}=0$ and $\phi_{\mathrm{trap}}=\pi$ in Eq.~\eqref{eq:Fluxonium_Hamiltonian2}. 
To precisely determine the value of $\phi_{\mathrm{trap}}$ when flux trapping was performed, we measured the qubit excitation spectrum around the sweet spot, as shown in the right panel of Fig.~\ref{fig2}(b).
By fitting the data points~(blue dots) to a parabolic curve~(black solid line), we determined the bias current of $2.35~\mathrm{\mu A}$ corresponding to the sweet spot. This indicated that our qubit was biased at $\phi_{\mathrm{trap}}/2\pi=0.50073$ in this particular flux-trapping operation. 

Next, we demonstrated flux trapping with various $I_{\mathrm{prebias}}$, the amount of current applied to the coil during the superconducting transition of the fluxonium circuit, and measured the resulting phase bias $\phi_{\mathrm{trap}}$. 
Here, we measured two flux-trapping fluxoniums in the same sample package as explained in Appendix~\ref{sec:A}. Assuming that the magnetic field generated by the coil was sufficiently homogeneous, we expected simultaneous flux trapping since they had the same ring radius.
Most of the data points were obtained from different cooling cycles of the dilution refrigerator. However, some data points were obtained by raising the sample-stage temperature above the superconducting transition temperature of the devices. The samples were then cooled down to cryogenic temperatures while maintaining a constant prebias current through the superconducting coil. We detail this procedure in Appendix~\ref{sec:B}.

Figure~\ref{fig3}(a) shows the resulting phase bias $\phi_{\mathrm{trap}}$ of the two flux-trapping fluxoniums. We observed binary phase biasing depending on $I_{\mathrm{prebias}}$ due to flux trapping. 
Within the range of $0.5\leq \Phi_{\mathrm{prebias}}/\Phi_0 \leq 1.5$, where a single flux quantum was expected to be trapped, two qubits were simultaneously biased at their sweet spots in most flux-trapping operations. We note that the flux-trapping operation with $I_{\mathrm{prebias}}=500~\mathrm{nA}$ was performed three times, all of which resulted in biasing around their sweet spots for both devices.

As we mentioned above, the ideal bias point is $\phi_{\mathrm{trap}} = 0~\mathrm{or}~\pi \pmod {2\pi}$ when the qubit and its surrounding structures are symmetric. 
Thus, we define the deviation of the phase bias from the theoretical bias point as
\begin{equation}
  \Delta\phi_{\mathrm{trap}}=
  \begin{cases*}
    \phi_{\mathrm{trap}} & ($\phi_{\mathrm{trap}}\simeq0$) \\
    \phi_{\mathrm{trap}}-\pi                 & ($\phi_{\mathrm{trap}}\simeq\pi$)
  \end{cases*}
\end{equation}
to discuss the asymmetry leading to the deviation.
Figure~\ref{fig3}(b) plots $\Delta\phi_{\mathrm{trap}}$, the deviation of phase bias $\phi_{\mathrm{trap}}$ from black dashed lines in Fig.~\ref{fig3}(a). Notably, most deviations observed in device 1 were within the range of $|\Delta\phi_{\mathrm{trap}}|<10^{-3}$ and did not display clear dependence on $I_{\mathrm{prebias}}$. Accordingly, the average and the variation were calculated to be $\Delta\phi_{\mathrm{trap}}/2\pi=(-0.1\pm0.9)\times10^{-3}$. As for device 2,
$\phi_{\mathrm{trap}}$ systematically deviated from the ideal values, and $\Delta\phi_{\mathrm{trap}}/2\pi$ was calculated to be $(-3.2\pm0.8)\times10^{-3}$. We mention possible reasons for these deviations in Sec.~\ref{sec:discussion}.

Additionally, we evaluated the temporal stability of this phase biasing. Figure~\ref{fig3}(c) shows $\phi_{\mathrm{trap}}$ of device 1 measured for a week after the flux-trapping operation, by which a single flux quantum was trapped with $I_{\mathrm{prebias}} = 500~\mathrm{nA}$. The phase bias consistently remained at $\phi_{\mathrm{trap}}=0.5+(9.0\pm0.1)\times 10^{-4}$, indicating temporally stable phase biasing achieved through flux trapping. We also note that phase jumps due to the escape of flux quanta from the ring, which were reported in previous work~\cite{Daria2022} in above-ground laboratory conditions, have not been observed in our experiments. The referenced work showed that flux-escape events at millikelvin temperatures are activated by ionizing radiation; none were detected in a deep-underground laboratory unless the devices had been exposed to an external source of the radiation. We attribute the longer flux-holding times of our device to the 20-$\mu$m-wide and 100-nm-thick TiN ring , which is robust enough against radiation-induced heating and quenching of the supercurrent. 

\begin{figure}[t]
\includegraphics[width=\columnwidth]{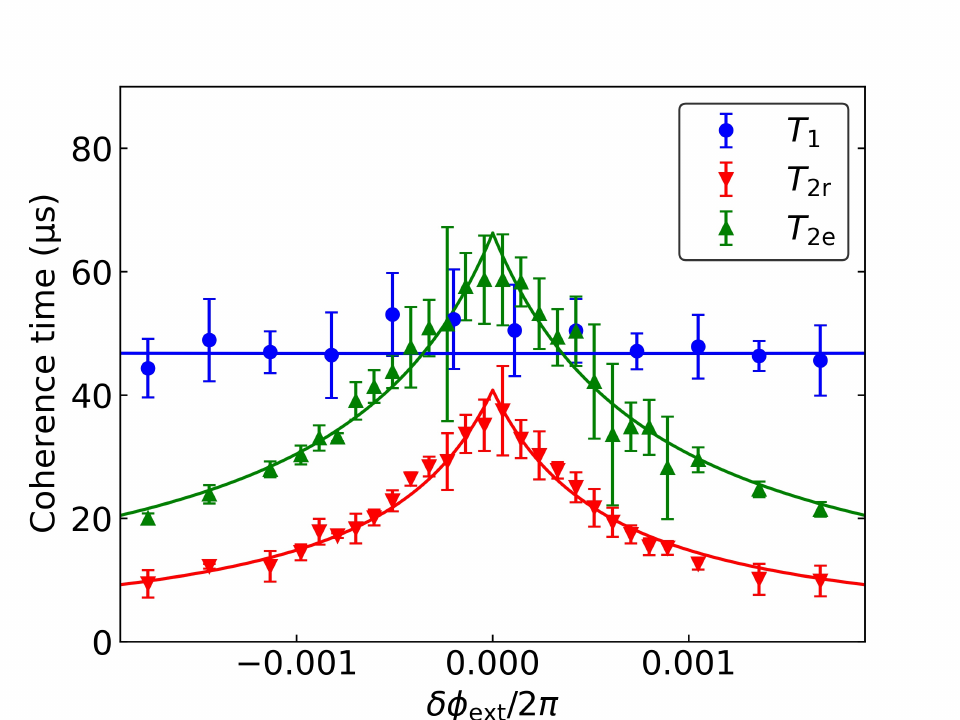}
\caption{
Coherence times of device 1 around its sweet spot when a single flux quantum is trapped. 
Measured $T_1$~(blue circle), $T_{\mathrm{2r}}$~(red inverted triangle), and $T_{\mathrm{2e}}$~(green triangle) are plotted with their fits to the model~(solid lines).
}
\label{fig4}
\end{figure}

For further characterization of our qubits, we measured the coherence times of device 1 around its sweet spot with a single flux quantum trapped. Figure~\ref{fig4} shows the dependence of coherence times on the flux bias offset from the sweet spot, $\delta\phi_{\mathrm{ext}}$.
First, we assume that the energy-relaxation time $T_1$ was limited by dielectric loss and obtained the dielectric loss tangent $\tan{\delta_{\mathrm{C}}}=2.0\times10^{-6}$ by fitting the measured $T_1$ values. This value is within the range of those measured in other fluxoniums~\cite{Nguyen2019, Somoroff2023}. 
Second, we assume that the dephasing time $T_2$ was limited by energy relaxation, 1/$f$ flux noise, and other noise independent of $\delta\phi_{\mathrm{ext}}$. By fitting measured Ramsey dephasing times $T_{\mathrm{2r}}$ and echo dephasing times $T_{\mathrm{2e}}$, we obtain the 1/$f$ flux noise amplitude at 1~Hz as $A_\Phi^{\mathrm{R}}=4.6~\mathrm{\mu\Phi_0/\sqrt{Hz}}$ and $A_\Phi^{\mathrm{e}}=6.6~\mathrm{\mu\Phi_0/\sqrt{Hz}}$, respectively. These values are slightly larger than the value reported in previous work~\cite{Nguyen2019}. We attribute these relatively large noise values to the dimensions of our qubit rings—both their size and width—which lead to increased fluctuations in threading magnetic flux. 
We further discuss these coherence times and those of device 2 in Appendix~\ref{sec:E} and conclude that the flux-trapping operation does not degrade the coherence property of the fluxonium.

\section{Discussion} \label{sec:discussion}

In this section, we discuss remaining issues and possible improvements in the phase biasing of the flux-trapping fluxonium.

We observed a few instances of unexpected phase biasing in Fig.~\ref{fig3}(a). For example, the data points at $\phi_{\mathrm{trap}}\simeq0$ for $I_{\mathrm{prebias}}=300$ and $800~\mathrm{nA}$ could be attributed to the small difference between the actual amount of applied magnetic flux and estimated $\Phi_{\mathrm{prebias}}$. However, the phase biasing of device 1 at $\phi_{\mathrm{trap}}\simeq\pi$ for $I_{\mathrm{prebias}}=0~\mathrm{nA}$ and of device 2 at $\phi_{\mathrm{trap}}\simeq0$ for $I_{\mathrm{prebias}}=600~\mathrm{nA}$ cannot be explained.
The number of trapped flux quanta is determined by the amount of magnetic flux threading through the superconducting ring during its superconducting transition, as described in Sec.~\ref{sec:principle}. Thus, the irregular flux-trapping instances indicate the presence of an unpredicted magnetic field during the superconducting transition or flux-escape events. 
We suspect that a thermal-gradient-induced current generates the stray magnetic field.
The flux-escape event can be caused by a large magnetic penetration depth $\lambda$ just below $T_{\mathrm{c}}$, leading to flux escape from ionizing radiation, as observed in the previous study~\cite{Daria2022}.
We expect that improved thermal anchoring and radiation shielding of the devices will mitigate these problems. 

Second, Fig.~\ref{fig3}(b) shows the remaining deviations from the ideal bias points. While the phase biasing of device 1 randomly deviated around $\Delta\phi_{\mathrm{trap}}=0$, device 2 showed systematic phase deviations, even for $I_{\mathrm{prebias}}=0~\mathrm{nA}$. This indicates the presence of an inhomogeneous ambient magnetic field that systematically phase-biases the qubit.
We also attribute the systematic deviations to asymmetric structures of the device caused by a fabrication error. We patterned the qubit structures made of a TiN film using a UV laser writer~(see the details in Appendix~\ref{sec:D}).
Considering the ring radius of the qubit is about $100~\mathrm{\mu m}$, the observed deviations of the order of $0.1\%$ correspond to a deviation of the order of $100~\mathrm{nm}$ from the design, which is well below the minimum resolution of the lithography tool. The possible asymmetric structure causes a difference in the amount of magnetic flux in the two sub-loops of the flux-trapping fluxonium generated by the associating supercurrent, which in turn phase-biases the gradiometric qubit. If this is the case, we can observe a clear dependence on $I_{\mathrm{prebias}}$, as the phase bias scales with the number of trapped flux quanta. To quantitatively evaluate the contribution, experiments trapping a larger number of flux quanta can amplify the deviation of the phase bias from the target, allowing for a more definitive assessment of this hypothesis.

Additionally, both devices 1 and 2 displayed random phase deviations of the same order of magnitude, which cannot be explained by fabrication errors and an ambient magnetic field.
These deviations were possibly caused by vortices randomly trapped in the defects in the superconducting film. Radially polarized spins on the surface of the superconducting film, which are known to give a phase offset~\cite{Sendelbach2008}, also account for the random deviations: At cryogenic temperatures, surface spins on the superconducting electrodes interact with each other, forming a magnetic order which phase-biases the qubit. 

We expect that several improvements could enhance the accuracy and precision of phase biasing: using e-beam lithography for higher-resolution patterning of qubit structures, carefully configuring moats~(as commonly used in SFQ circuits~\cite{Bermon1983, Yamanashi2018}), improving magnetic shielding, and employing wider qubit rings to suppress surface-spin effects~\cite{Bialczak2007, Braumuller2020}.

\section{Conclusions} \label{sec:conclusion}
In this work, we developed a flux-trapping fluxonium, which enables biasing the qubit at its sweet spot by utilizing fluxoid quantization in a flux-trapping ring, and achieved accurate phase biasing.
Additionally, we ensured the temporal stability of phase biasing for at least a week, achieved simultaneous phase biasing of two qubits with a single superconducting coil, and confirmed that the coherence property was not degraded by the flux-trapping operation.
With the phase biasing using flux trapping, we no longer need a large current to bias a fluxonium. This allows us to reduce the product of the mutual inductance between a flux bias line and the fluxonium and the bias current by orders of magnitude, mitigating decoherence and crosstalk.
With further improvements we have discussed, flux trapping would eliminate the need for external magnetic flux bias lines for individual qubits and mitigate problems caused by introducing those bias lines, paving the way for fluxonium integration. 
Furthermore, our flux-trapping protocol is applicable to other types of superconducting qubits that require external magnetic flux bias, facilitating their scaling to large superconducting quantum processors.
\begin{acknowledgments}
We gratefully acknowledge K. Kato and T. Miyamura for their insightful discussions, and H. Terai and Y. Hishida for their meticulous preparation of the TiN-deposited wafer. M. F\'echant, D. B\'en\^atre and I. Pop kindly gave us their valuable comments and suggestions, particularly regarding the radiation-induced flux escapes observed in their previous work.
This work was supported in part by JST ERATO~(Grant No.~JPMJER1601), MEXT QLEAP~(Grant No.~JPMXS0118068682), and JST CREST~(Grant No.~JPMJCR23I4).

\end{acknowledgments}

\begin{appendices}
\renewcommand{\thefigure}{A\arabic{figure}}
\renewcommand{\thetable}{A\arabic{table}}
\renewcommand{\theHfigure}{A\arabic{figure}}
\renewcommand{\theHtable}{A\arabic{figure}}
\setcounter{figure}{0}
\setcounter{table}{0}

\section{EXPERIMENTAL SETUP}\label{sec:A}

\begin{figure}[t]
\includegraphics{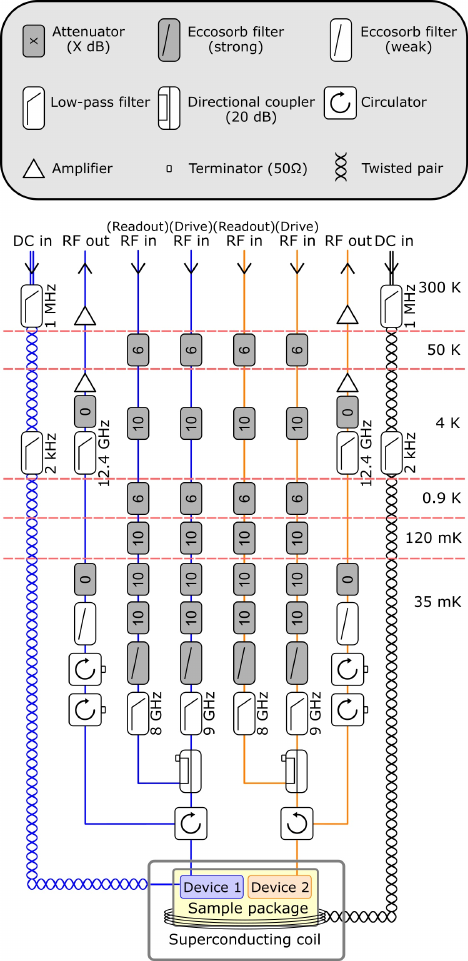}
\caption{Schematic diagram of the wiring inside the dilution refrigerator.}
\label{figS1}
\end{figure}

Our experiments were conducted in a dilution refrigerator~(Oxford Triton200). Figure~\ref{figS1} represents the experimental setup inside. Two devices are installed in the same sample package and enclosed by a magnetic shield.
A superconducting coil is located inside the magnetic shield, surrounding the sample package, which enables simultaneous flux-trapping operations for the two devices. The parameters of these two devices are shown in Table~\ref{tab1}. Each RF-input line is attenuated at each temperature stage, totaling 52 dB of attenuation. In addition, a low-pass filter and an Eccosorb filter are added at the lowest-temperature stage to suppress thermal noise from higher-temperature stages. Two RF-input lines, for qubit drive and readout tones respectively, are combined with a directional coupler and connected to a resonator port of each device. Reflected signals are transmitted through three circulators, an Eccosorb filter with weaker attenuation than that of the one on the input lines, and a low-pass filter on an RF-output line. The two 0-dB attenuators on each line work as thermal contacts with the respective stages. Subsequently, the signals are amplified with a high-electron-mobility-transistor~(HEMT) amplifier at the 4-K stage. A DC current is applied through a twisted-pair line with a low-pass filter with a cut-off frequency of 2~kHz at the 4-K stage to the superconducting coil or an on-chip bias line of device 1. The superconducting coil is also used to phase-bias device 2 at the operation temperature, which is enabled by the slight asymmetry in the device structure, the inhomogeneity of the magnetic field generated by the coil, or both.
At room temperature, typically 20--30-dB attenuators are added to the RF-input lines, and reflected signals are further amplified with a low-noise amplifier. There is also a $\pi$-filter with a cut-off frequency of 1~MHz for each DC line.

\begin{table}[t]
\caption{Qubit and resonator parameters in devices~1 and~2. $E_{\mathrm{J}}$, $E_{\mathrm{C}}$ and $E_{\mathrm{L}}$ are the energy scales describing the fluxonium Hamiltonian. $\omega_{\mathrm{r}}$ is the readout-resonator frequency, $\chi_{01}$ is the dispersive shift, and $\kappa_{\mathrm{r}}$ is the resonator linewidth.}
\label{tab1}
\centering
\begin{tabular}{lrrrrrr}
\hline
\hline
 & ~~$E_{\mathrm{J}}/h$ & ~~$E_{\mathrm{C}}/h$ & ~~$E_{\mathrm{L}}/h$ & ~~$\omega_{\mathrm{r}}/2\pi$ &~$\chi_{01}/2\pi$ & ~$\kappa_{\mathrm{r}}/2\pi$\\ 
  & (GHz) & (GHz) & (GHz)  & (GHz)& (MHz) & (MHz)\\ \hline
 Device 1& 3.54 & 1.32 & 0.81 & 7.988 & 0.4 & 10.8 \\ 
 Device 2& 3.52 & 1.31 & 0.66 & 7.405 & 0.7 & 11.4 \\ \hline\hline
\end{tabular}
\end{table}

\section{FLUX-TRAPPING OPERATION}\label{sec:B}
\renewcommand{\thefigure}{B\arabic{figure}}
\renewcommand{\thetable}{B\arabic{table}}
\renewcommand{\theHfigure}{B\arabic{figure}}
\renewcommand{\theHtable}{B\arabic{table}}
\setcounter{figure}{0}
\setcounter{table}{0}

In the experiments shown in Fig.~\ref{fig3}, several flux-trapping operations were conducted successively without warming up the dilution refrigerator to room temperature. During this process, the outer ring of the flux-trapping fluxonium needs to transition from the normal conducting phase to the superconducting phase with a prebias current applied to the coil.

The procedure goes as follows: First, we start collecting the mixture of $^3$He/$^4$He circulating inside the dilution refrigerator. The temperature at the sample stage, where the device is installed, rises due to the heat conduction from the upper stages. Throughout this process, we monitor the temperature using a $\mathrm{RuO}_2$ thermometer mounted at the sample stage.
After confirming the superconducting-to-normal phase transition as described below, we stop collecting the mixture. Then, a constant current for the flux trapping is applied to the coil, and we cool the dilution refrigerator to the lowest temperature before switching off the current.

\begin{figure}[t]
\includegraphics{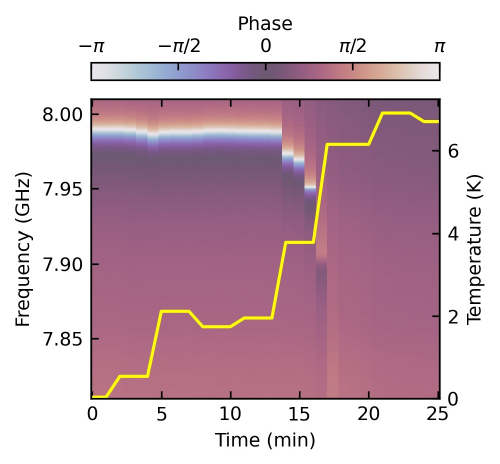}
\caption{Time evolutions of the resonator response and the mixing-chamber-plate temperature during the warming-up of the dilution refrigerator. The color plot shows the reflection phase of the resonator response. The yellow solid line represents the temperature measured by the thermometer installed at the mixing-chamber plate. The time on the horizontal axis is measured from the start of the warming-up process.}
\label{figS2}
\end{figure}

Figure~\ref{figS2} shows an example of time evolutions of the resonator response of device 1 and the sample-stage temperature, after starting to warm up the dilution refrigerator.
Both the superconducting ring of the flux-trapping fluxonium and the readout resonator are made of a TiN thin film. Its superconducting transition temperature $T_{\mathrm{c}}$ in the bulk state is approximately $5.6~\mathrm{K}$, and the thermometer displayed a temperature exceeding $T_{\mathrm{c}}$ after 17 minutes. The resonance frequency and Q value of the resonator sharply dropped just before this point.
We attribute this behavior to the transition from the superconducting phase to the normal conducting phase. Afterward, the thermometer showed a temperature above $7~\mathrm{K}$, and the resonator response was no longer visible, indicating that the temperature of the flux-trapping fluxonium exceeded $T_{\mathrm{c}}$. At this point, we applied the prebias current to the coil for the flux-trapping operation and started cooling the dilution refrigerator.

\section{CALIBRATION OF A SUPERCONDUCTING COIL}\label{sec:C}
\renewcommand{\thefigure}{C\arabic{figure}}
\renewcommand{\thetable}{C\arabic{table}}
\renewcommand{\theHfigure}{C\arabic{figure}}
\renewcommand{\theHtable}{C\arabic{table}}
\setcounter{figure}{0}
\setcounter{table}{0}

\begin{figure}[t]
\includegraphics{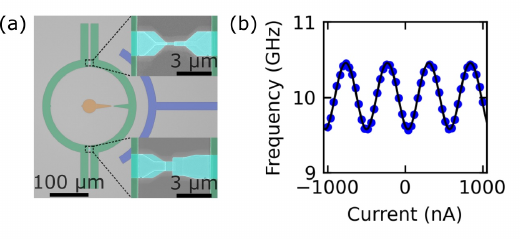}
\caption{Calibration of the magnetic flux by using a SQUID-transmon qubit with the same ring size as the flux-trapping fluxonium.
(a) False-colored image of the SQUID-transmon device. A tunable transmon qubit~(green) is capacitively coupled to a resonator pad~(blue). The ring radius is $100~\mathrm{\mu m}$. An inner pad~(orange) is not connected to the capacitance pads of the transmon but is fabricated to mimic the structure of the flux-trapping fluxoniums. Two small Josephson junctions with different areas~(light blue) are fabricated across the gaps along the ring. 
(b) Excitation frequency of the tunable transmon qubit~(blue dots) as a function of the current applied to the superconducting coil. The black solid line represents a fit to Eq.~(\ref{eq:transmon_freq}), with $E_{\mathrm{J1}}/h = 39.0~\mathrm{GHz}$, $E_{\mathrm{J2}}/h = 3.3~\mathrm{GHz}$, and $E_{\mathrm{C}}/h = 0.35~\mathrm{GHz}$.}
\label{figS3}
\end{figure}

We calibrated the magnetic flux threading the rings of our qubits, which was generated by the superconducting coil, using a SQUID-transmon qubit with the same ring radius~[Fig.~\ref{figS3}(a)]. 
The Hamiltonian of a SQUID-transmon qubit is described as

\begin{equation}\label{eq:Transmon_Hamiltonian}
\hat{\mathcal{H}}
=4E_{\mathrm{C}}(\hat{n}-n_\mathrm{g})^2-E_{\mathrm{J1}}\cos{\hat{\phi}}-E_{\mathrm{J2}}\cos{(\hat{\phi}-\phi_{\mathrm{ext}})},
\end{equation}
where $n_\mathrm{g}$ is the offset charge, $E_{\mathrm{J1}}$ and $E_{\mathrm{J2}}$ are the Josephson energies of the Josephson junctions, $\phi_{\mathrm{ext}}=2\pi \Phi_{\mathrm{ext}}/\Phi_0$ is the phase offset when the external magnetic flux $\Phi_{\mathrm{ext}}$ threads the SQUID ring, and others follow a similar definition as given in Eq.~(\ref{eq:Fluxonium_Hamiltonian}) in the main text.
Assuming $E_{\mathrm{C}}\ll |E_{\mathrm{J1}}-E_{\mathrm{J2}}|$, the transition frequency of this qubit is approximately calculated as
\begin{equation}\label{eq:transmon_freq}
E_{01}
=\sqrt{8E_{\mathrm{C}}E_{\mathrm{J}}(\phi_{\mathrm{ext}})}-E_{\mathrm{C}},
\end{equation}
with $E_{\mathrm{J}}(\phi_{\mathrm{ext}}) \coloneqq \sqrt{E_{\mathrm{J1}}^2+E_{\mathrm{J2}}^2+2E_{\mathrm{J1}}E_{\mathrm{J2}}\cos{\phi_{\mathrm{ext}}}}$. We see the transition frequency as a function of the external magnetic flux $\Phi_{\mathrm{ext}}$ is periodic with a period of $\Phi_0$. Therefore, the amount of current required to apply a single flux quantum inside the ring can be extracted by measuring the modulation period of the qubit frequency as a function of the current through the coil.

Figure~\ref{figS3}(b) shows the spectrum of the SQUID-transmon qubit as a function of the current through the coil. From the periodicity of the spectrum, we determined that a current of $540~\mathrm{nA}$ generates approximately a single flux quantum threading the ring of our qubits.
Thus, in the absence of background magnetic fields, the magnetic flux threading the ring of the flux-trapping fluxonium, $\Phi_{\mathrm{prebias}}$, generated by the current through the coil, $I_{\mathrm{prebias}}$, can be calculated as $\Phi_{\mathrm{prebias}}/\Phi_0 = I_{\mathrm{prebias}}/(540\,\mathrm{nA})$.

\section{DEVICE FABRICATION}\label{sec:D}
\renewcommand{\thefigure}{D\arabic{figure}}
\renewcommand{\thetable}{D\arabic{table}}
\renewcommand{\theHfigure}{D\arabic{figure}}
\renewcommand{\theHtable}{D\arabic{table}}
\setcounter{figure}{0}
\setcounter{table}{0}
The devices used in this work were fabricated on a 300-$\mu$m-thick Si wafer with a 100-nm-thick TiN film deposited on it.

First, the device structures, except for the Josephson junctions, were drawn using a UV laser writer~(Heidelberg Maskless Aligner MLA 150) after coating with positive photoresist~(AZ1500). This was followed by the development and reactive ion etching with CF$_4$ for patterning. After removing the resist with NMP-DMSO solution, the wafer was cleaned with O$_2$ plasma ashing and BHF solution.

Second, the junction masks were fabricated with the double layers of positive electron-beam resist, MMA-MAA Copolymer EL-11 and 495 PMMA A4, exposed with an e-beam writer~(Raith EBPG5150). After development in IPA-DI solution and O$_2$ plasma descum, Al was deposited in a Plassys electron-beam evaporator with double-angle evaporation.
The wafer was subsequently cleaned with NMP-DMSO solution and diced into $2.5\times5~\mathrm{mm^2}$ chips.

\section{COHERENCE CHARACTERIZATION}\label{sec:E}
\renewcommand{\thefigure}{E\arabic{figure}}
\renewcommand{\thetable}{E\arabic{table}}
\renewcommand{\theHfigure}{E\arabic{figure}}
\renewcommand{\theHtable}{E\arabic{table}}
\setcounter{figure}{0}
\setcounter{table}{0}
We detail the analysis of the coherence times of device~1 around its sweet spot shown in Fig.~\ref{fig4}.

First, we obtain the relaxation time $T_1$ by fitting the decay of the excited-state population $P_{\mathrm{e}}$ to $P_{\mathrm{e}}=Ae^{-\Gamma_1 t}+B$, where $\Gamma_1=1/T_1$ is the relaxation rate. 
We assume that the relaxation time is mainly limited by dielectric loss, as often seen in previous works~\cite{Nguyen2019, Helin2021}, and model the relaxation rate as
\begin{equation}\label{eq:T1_loss}
\Gamma_1=
\frac{16E_{\mathrm{C}} \tan{\delta_{\mathrm{C}}}}{\hbar}|\braket{0|\hat{n}|1}|^2\coth{\frac{\hbar\omega_{01}}{2k_{\mathrm{B}} T}}.
\end{equation}
Here, $\tan{\delta_{\mathrm{C}}}$ is the dielectric loss tangent, $\omega_{01}$ is the qubit frequency, and $T$ is the temperature. 
By taking $T=50~\mathrm{mK}$ based on the measured excited-state population $P_{\mathrm{e}}=0.30$ at the equilibrium, we obtain the fitting curve as shown in Fig.~\ref{fig4} and $\tan{\delta_{\mathrm{C}}}=2.0\times10^{-6}$ at the qubit frequency, which is comparable to the values reported in previous works~\cite{Nguyen2019, Helin2021}.

Next, we measured the echo decay time $T_{\mathrm{2e}}$.
At the sweet spot, $T_{\mathrm{2e}}$ can be extracted by fitting $P_{\mathrm{e}}$ to $P_{\mathrm{e}}=Ae^{-\Gamma_{\mathrm{2e}} t}+B$, where $\Gamma_{\mathrm{2e}}=1/T_{\mathrm{2e}}$ is the echo decay rate. On the other hand, away from the sweet spot, the fluxonium becomes susceptible to 1/$f$ flux noise, leading to a Gaussian decay curve of the echo signal. Following the discussion in the previous work~\cite{Helin2021}, we fit $P_{\mathrm{e}}$ to 
$P_{\mathrm{e}}=Ae^{-\Gamma_{\mathrm{2,exp}} t-(\Gamma_{\mathrm{2,Gauss}} t)^2}+B$, where $\Gamma_{\mathrm{2,exp}}$ is the exponential decay rate and $\Gamma_{\mathrm{2,Gauss}}$ is the Gaussian decay rate,
and define $T_{\mathrm{2e}}$ as 
\begin{equation}
T_{\mathrm{2e}}=\frac{\sqrt{\Gamma_{\mathrm{2,exp}}^2+4\Gamma_{\mathrm{2,Gauss}}^2}-\Gamma_{\mathrm{2,exp}}}{2\Gamma_{\mathrm{2,Gauss}}^2}.
\end{equation}
This echo decay time $T_{\mathrm{2e}}$ is a characteristic time for the echo amplitude to decay by $1/e$ times.
We model the echo decay rate as
\begin{equation}
\Gamma_{\mathrm{2e}}=A_\Phi^{\mathrm{e}} \left|\frac{\partial\omega_{01}}{\partial\Phi_{\mathrm{ext}}} \right| \sqrt{\ln{2}}+\frac{\Gamma_1}{2}+\Gamma_{\phi}^{\mathrm{e,misc}},
\end{equation} 
with $A_\Phi^{\mathrm{e}}$ being the noise amplitude of $1/f$ flux noise at 1~Hz, and $\Gamma_{\phi}^{\mathrm{e,misc}}$  the dephasing rate caused by other types of noise independent of external flux. 
As a result, we obtain 
$A_\Phi^{\mathrm{e}}=6.6~\mu\Phi_0/\sqrt{\mathrm{Hz}}$ and $\Gamma_{\phi}^{\mathrm{e,misc}}/2\pi=4.4~\mathrm{kHz}$ by fitting the measured echo coherence times as shown in Fig.~\ref{fig4}.

\begin{figure}[t]
\includegraphics[width=\columnwidth]{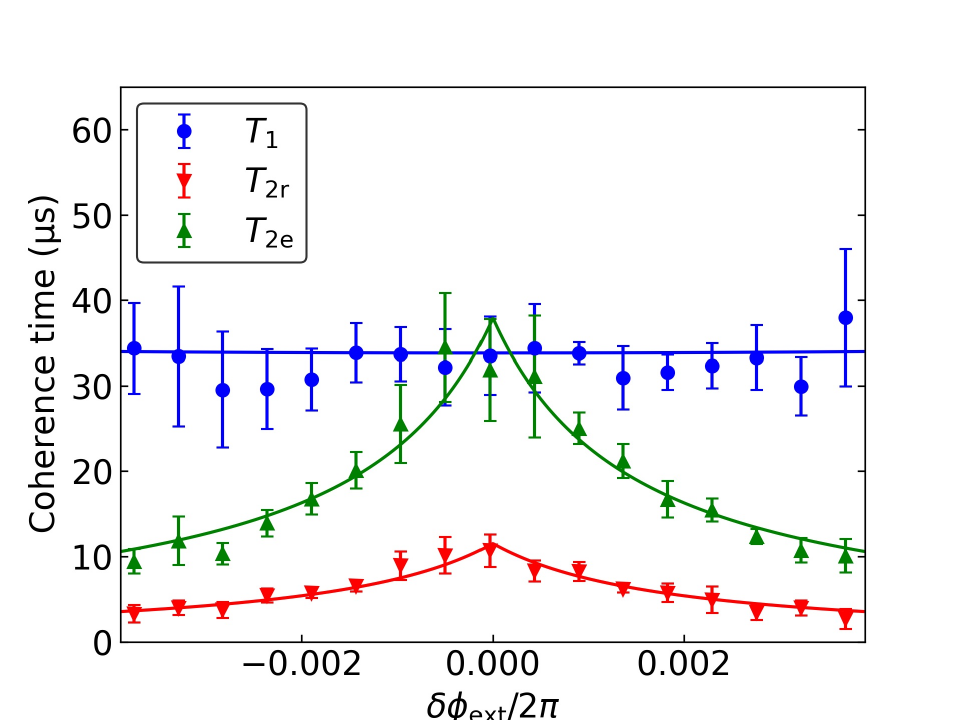}
\caption{Coherence times of device 2 around its sweet spot when a single flux quantum is trapped.}
\label{figS4}
\end{figure}

Likewise, we evaluate the Ramsey decay time
by fitting the decay curve to $P_{\mathrm{e}}=Ae^{-\Gamma_{\mathrm{2,exp}} t-(\Gamma_{\mathrm{2,Gauss}} t)^2}\cos{(\Delta \omega t+\varphi)}+B$, where $\Delta \omega$ is the detuning between the qubit frequency and the drive frequency. We model the Ramsey decay rate as
\begin{equation}
\Gamma_{\mathrm{2r}}=A_\Phi^{\mathrm{r}} \left|\frac{\partial\omega_{01}}{\partial\Phi_{\mathrm{ext}}} \right| \sqrt{\ln{\frac{\Gamma_{\mathrm{2r}}}{\omega_{\mathrm{l}}}}}+\frac{\Gamma_1}{2}+\Gamma_{\phi}^{\mathrm{r,misc}},
\end{equation}
where $A_\Phi^{\mathrm{r}}$ and $\Gamma_{\phi}^{\mathrm{r,misc}}$ respectively follow the definitions of $A_\Phi^{\mathrm{e}}$ and $\Gamma_{\phi}^{\mathrm{e,misc}}$, and $\omega_{\mathrm{l}}$ represents the infrared cut-off frequency of $1/f$ flux noise, which we set to $\omega_{\mathrm{l}}/2\pi=1~\mathrm{Hz}$ based on our measurement protocol.
The resulting parameters are
$A_\Phi^{\mathrm{r}}=4.6~\mu\Phi_0/\sqrt{\mathrm{Hz}}$ and $\Gamma_{\phi}^{\mathrm{r,misc}}/2\pi=14~\mathrm{kHz}$.
As mentioned in the main text, $A_\Phi^{\mathrm{e}}$ is larger than $A_\Phi^{\mathrm{r}}$. This indicates that the flux-noise spectrum deviated from the 1/$f$ spectrum, as also discussed in previous work~\cite{Feng2022}. 
Additionally, $\Gamma_{\phi}^{\mathrm{e,misc}}$ was smaller than $\Gamma_{\phi}^{\mathrm{r,misc}}$, suggesting the presence of low-frequency noise at the sweet spot.
Photon-shot noise in a readout resonator often accounts for dephasing of a fluxonium at the sweet spot. However, in our device, it is calculated to be on the order of $10~\mathrm{Hz}$ because of the small $\chi_{01}/\kappa_{\mathrm{tot}}$ ratio, as shown in Table~\ref{tab1}. Thus, we suspect that the qubit was dephased by fluctuations in the critical current of the Josephson junction~\cite{Harlingen2004}. 

We also measured the coherence times of device 2 when a single flux quantum was trapped through the flux-trapping operation~(Fig.~\ref{figS4}). From the fittings, we obtain the noise parameters; $\tan{\delta_{\mathrm{C}}}=3.2\times10^{-6}$, $A_\Phi^{\mathrm{e}}=7.5~\mu\Phi_0/\sqrt{\mathrm{Hz}}$, $\Gamma_{\phi}^{\mathrm{e,misc}}/2\pi=12~\mathrm{kHz}$,
$A_\Phi^{\mathrm{r}}=5.5~\mu\Phi_0/\sqrt{\mathrm{Hz}}$, and $\Gamma_{\phi}^{\mathrm{r,misc}}/2\pi=72~\mathrm{kHz}$.
The coherence times were slightly shorter than those of device~1 but can be modeled in the same manner.

\end{appendices}

\bibliography{main}

\end{document}